\documentclass[graybox,envcountchap,sectrefs]{usb_TEX}

\usepackage{hyperref}
\hypersetup{
    pdfauthor={Wilder Castellanos},
    pdftitle={Internet of things: a multiprotocol gateway as solution of the interoperability problem},
    colorlinks,
    citecolor=blue,
    filecolor=black,
    linkcolor=blue,
    urlcolor=blue
}
\usepackage{subfigure}
\usepackage{mathptmx}
\usepackage{helvet}
\usepackage{courier}
\usepackage{apacite}
\usepackage{type1cm}         
\usepackage{makeidx}         
\usepackage{graphicx}        
\graphicspath{{Figures/}}
\usepackage{multicol}        
\usepackage[bottom]{footmisc}
\usepackage{iftex}
\usepackage{amsmath}
\usepackage{amssymb}
\usepackage{captdef}
\usepackage{chemformula}
\usepackage{multirow}

\makeindex             

\newcommand*{\doi}[1]{\href{#1}{#1}}

\newcommand{\chapterauthor}[1]{%
  {\vspace*{-70pt}%
   \linespread{1.1}\large\scshape#1%
  \vspace*{35pt}}
}
\makeatother

\begin{document}

\mainmatter

%
%
%
\chapter*{Chapter 5.\\ 
Internet of things: a multiprotocol gateway as solution of the interoperability problem}

\chapterauthor{Wilder Castellanos
\footnote{Programa de Ingenier\'ia Electr\'onica, Universidad de San Buenaventura sede Bogot\'a, Cra. 8H No. 172-20, Bogot\'a, Colombia. e-mail: wcastellanos@usbbog.edu.co} 
Jose Macias
\footnote{Programa de Ingenier\'ia Electr\'onica, Universidad de San Buenaventura sede Bogot\'a, Cra. 8H No. 172-20, Bogot\'a, Colombia. e-mail: jmacias@academia.usbbog.edu.co}, 
Harold Pinilla 
\footnote{Programa de Ingenier\'ia Electr\'onica, Universidad de San Buenaventura sede Bogot\'a, Cra. 8H No. 172-20, Bogot\'a, Colombia. e-mail: hpinilla@academia.usbbog.edu.co}, 
Jose David Alvarado
\footnote{Programa de Ingenier\'ia Electr\'onica, Universidad de San Buenaventura sede Bogot\'a, Cra. 8H No. 172-20, Bogot\'a, Colombia. e-mail: jalvarado@usbbog.edu.co}.
}


\label{Chapter_5} 

\abstract{In the coming years, the interconnection of a large number of devices is expected, which will lead to a new form of interaction between the real and the virtual world. In this promising scenario, known as the Internet of Things (IoT), it is expected that different objects, such as sensors, industrial robots, cars, appliances, etc., will be continuously connected to the Internet. One of the main challenges of the Internet of Things is the interoperability of highly heterogeneous devices, mainly in terms of the communication capabilities and network protocols used. As consequence, the interconnection model of the different devices involves an intermediary device, known as gateway. This gateway is a centralized element for the management of the devices that make up an IoT application. In addition, it is essential for the transmission of information to the Internet, especially when many IoT devices are not IP-based.
This chapter describes a proposed model for an IoT gateway that allows the exchange of data through different wireless technologies and forwarding of such data to the Internet. The proposed gateway has important advantages such as: supporting for multiprotocol interconnectivity; the remote configuration of wireless nodes for sensor and actuators management; a flexible algorithm to translate the data obtained by sensors into a uniform format for transmission to a cloud server; low energy consumption due to efficient data transfer over the MQTT protocol. In order to demonstrate the usefulness of the developed gateway, a proof-of-concept test was implemented. The implemented scenario consists of 2 wireless nodes responsible for sensing environmental variables and transmitting data to the gateway node through different communication protocols. The obtained results show the feasibility for simultaneous data transmission from the remote wireless nodes to the gateway. Metrics on energy consumption in the devices are also presented.}
\textit{Keywords:} Internet of Things, Wireless Sensor Network, Multiprotocol gateway, MQTT.

\section{The Internet of Things: perspectives and challenges}
\label{sec:5.1}

\begin{figure}[ht!]
\centering
\includegraphics[scale=1.0]{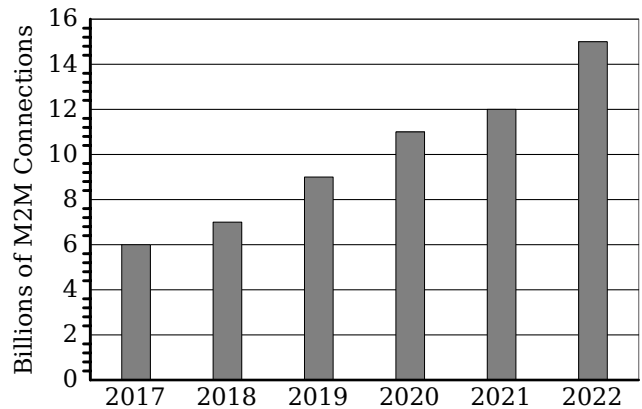}
\caption{Number of M2M connections per year}
\label{ch5_fig:1}       
\end{figure}

The Internet of Things is a new technological paradigm that promotes the vision of a global network of devices capable of interacting with each other \cite{Gubbi2013,Lee2015}. In IoT scenarios these all devices acquire information, process it and share it. Current studies on the ecosystem associated with IoT estimate 127 new IoT devices connect to the Internet every second \cite{Lucero2016}. This implies that in 2022 the M2M (Machine-to-Machine) traffic could reach 25ExaBytes per month (7\% of all Global internet traffic) \cite{Cisco2019}. This significant increase in traffic between devices will be due to the significant growth in the number of machines connected. Cisco estimates that M2M connections will grow from 6.1 Billion in 2017 to 14.6 billion by 2022 (Fig \ref{ch5_fig:1}). Connected home applications such as connected household appliances, video surveillance, home automation and tracking applications, will represent 48 \%, of the total M2M connections by 2022 \cite{Cisco2019}. 
%

From the point of view of the economy, the effect of the IoT incursion will also be considerable. In this sense, in the report titled “Global Internet of Things (IoT) market: Global market analysis, insights and forecast, 2019-2026” \cite{FBI2019} the global market of was valued at US\$ 190 Billion in 2018 and is estimated to reach \$ 1111.3 Billion by 2026 \cite{USAIDCC2019}. Some reports also predict the global market of IoT will expand at a CAGR (Compound annual growth rate) of 24.7 \% during the next 5 years. For example, the market for wearable devices is expected to grow to 162.9 million devices by the end of 2020, mainly led by Fitness trackers. Smart homes are also an important area in IoT, according to Business Insider forecasts (Meola, 2018), they will grow from 82 million in 2015 to 193 million in 2020 this will represent almost US\$58.7 Billion in 2020 \cite{Statista2018}. In transportation area, will In the area of transport, 220 million connected cars are estimated in 2020, approximately 190 million of cars more than in 2015 \cite{Meola2018}. In the agriculture sector, growth in the number of IoT devices is estimated from 30 million in 2015 to 75 million in 2020. In Banking and Financial Services market, the IoT will grow from US\$0.17 Billion in 2017 to 2.03 Billion by 2023. These forecasts show a new mode of interaction with the physical world, inspired by the idea of ubiquity, where all the devices that surround us (sensors, cars, refrigerators, thermostats, industrial robots, tables, smartphones, etc.) can be connected to Internet anytime and anywhere. 

However, several challenges must be overcome. Mainly those related to security and interoperability issues of the devices. Interoperability is not a minor issue due to the high degree of heterogeneity of the available IoT devices. This heterogeneity is mainly accentuated in communication capabilities (protocols, technologies and hardware) of the IoT components. Some figures that illustrate the magnitude of the problem are provided by some preliminary studies \cite{Alvarado2018, alvarado_2017}. For example, about 10 different communication standards for IoT systems have been recently identified. Some IoT protocols are: Bluetooth, WiFi, ZigBee, 6LoWPAN, LoRa and Sig-Fox. This implies a difficulty during the implementation of a real system where it is necessary to guarantee full interoperability. In particular, the interoperability requirement is one of the major challenges that must be addressed for the integration and development of new IoT platforms \cite{Al-Fuqaha2015}. Some studies estimates that about 40\% of the potential benefits of IoT could be lost due to the lack of interoperability between different devices \cite{Manyika2015}. Solving the interoperability problem will eliminate so-called closed ecosystems \cite{Desai2015}, obtaining the true value of IoT, that is, data that is acquired and transmitted through the interaction between devices \cite{Evans2011}. One possible solution is the integration of a multiprotocol IoT gateway capable of performing data exchange with several wireless nodes, through different network protocols such as WiFi, Bluetooth and Zigbee. An IoT gateway also establishes a connection to a server on the Internet to perform data analytics. 

In order to show how an IoT gateway should be implemented, we proposed a gateway architecture that enables data communication using different network protocols. The evaluation of the proposed gateway was carried out through a use case, which consisted of the implementation of two wireless nodes, each with 6 sensors. Data readings are transmitted from wireless nodes to gateway through WiFi, Bluetooth and Zigbee. Gateway performance was analyzed based on the measurement of the percentage of CPU usage and free memory, as well as the analysis of the throughput reached in each network interface.	

\section{IoT Gateway: bridge device between sensor network and cloud platforms}
\label{sec:5.2}

In general terms, a gateway is a device that allows communication between networks with different protocols. This task required to convert the formats of the received data to the structure of the destination protocol. In the scenario of IoT, a gateway is a device that supports communications with different communication protocols and data formats. Therefore, an IoT gateway allows to interconnect multiple types of sensors as well as the aggregation of several IoT nodes with other segments of the network or with Internet. This is the main objective of a gateway, to serve as a bridge between several network domains with a public network or the Internet, solving the problem of heterogeneity between these domains \cite{Zachariah2015}. At the same time, a gateway becomes an ideal device for network management functions, since while exchanging messages with the sensor nodes, it can map the network and establish a comprehensive knowledge of the network. 

\begin{figure}[ht!]
\centering
\includegraphics[scale=1.0]{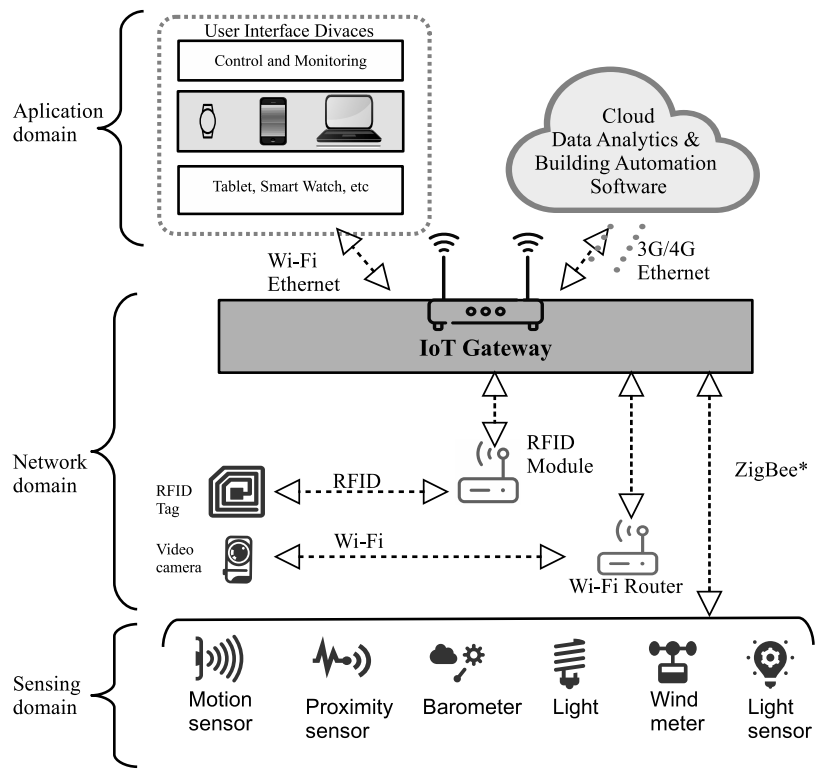}
\caption{Main characteristics for an IoT Gateway}
\label{ch5_fig:2}       
\end{figure}

The different types of gateways for IoT can be classified as passive, semi-automatic and fully automatic. In passive gateways new IoT devices, as sensor nodes, must be configured manually. Likewise, old devices that are no longer part of the network must be removed. An example of this type gateways is the one presented in reference \cite{Emara2009}. On the other hand, semi-automatic gateways can generate a link with the new IoT node but cannot automatically support the establishment of all configuration parameters \cite{Wu2013}. Finally, gateways that are fully automatic allow to self-configure new IoT devices and thus quickly solve heterogeneity problems during data transmission \cite{Kang2017}.

Next, some gateways that have been developed to improve interconnectivity between IoT devices will be described. For example, Gloria et al, in reference \cite{Gloria2017} describe the implementation of an IoT gateway that is the main component of a system dedicated to the monitoring and remote control of a pool. The gateway was developed using a Raspberry Pi board with full-duplex communication for data exchange between users and sensor networks. The sensors used in this platform are linked to an Arduino board which provides the connectivity to the gateway. The implemented sensors were: temperature, humidity, luminosity and level of water. However, data communication between the sensor network and the gateway is done through a USB cable. Thus, it is not feasible platform to implement in a real environment, due to the dependence of a wired connection.

Another gateway is the one proposed in \cite{Kang2018}, which is based on the IoTivity platform \cite{LinuxFoundation2019}. This gateway is focused on ensuring interoperability between devices with non-IP-based communication capabilities. To fulfill this objective, this gateway has implemented the CoAP (Constrained Application Protocol) protocol \cite{Shelby2014}. One of the advantages of this gateway is the self-configuration option, which allows a new sensor node to be connected to establish a dialogue with the gateway in order to be recognized as a new member of the system.

Other IoT gateways have been developed with the ESP32 card. For example, the implemented platform described in \cite{Khanchuea2019}, establishes a mesh network to data exchange between gateway and a sensor network. The sensor nodes use the DHT22 sensor to record temperature and humidity values. These nodes establish a ZigBee network to send the sensed data to a router node. This router node takes the data received by ZigBee link and relays it over WiFi to the gateway node. Although in this system there is a co-existence of two wireless networks (ZigBee and WiFi), the gateway node only establishes WiFi connections, so it lacks the property of being multiprotocol. 

Examples of multiprotocol gateways for IoT, can be consulted in references \cite{Guoqiang2013,YacchiremaVargas2016}. These gateways have as main function the conversion of protocols mainly between Zigbee, Bluetooth, WiFi and Ethernet. In these gateways data is stored and displayed through a web server embedded within the same gateway, which is not entirely practical if the possibility of a real application is analyzed. The most suitable would be the use of a cloud hosting service. 

According to the needs of the IoT ecosystem and the current needs of the IoT applications, we consider that an IoT gateway must have mainly the following characteristics:

\begin{itemize}
	\item Energy autonomy
	\item Multiprotocol
	\item Connectivity to Cloud Services
	\item Self-configuring
\end{itemize}

In next sections, we present an example of an IoT Gateway where some of the above features have been implemented. Also, we present a use case of the implemented gateway where a weather station with two wireless nodes transmit environmental data to the gateway. This data exchange between sensors nodes and gateway is simultaneously performed through three different communication protocols. Then, the IoT gateway delivers environmental data to a server in the Internet.

\section{A multiprotocol gateway for IoT: a proposed solution for the interoperability}
\label{sec:5.3}

A description of the architecture, hardware and software components of the developed IoT gateway will be given below. 
\subsection{Gateway architecture}

\begin{figure}[ht!]
\centering
\includegraphics[scale=0.8]{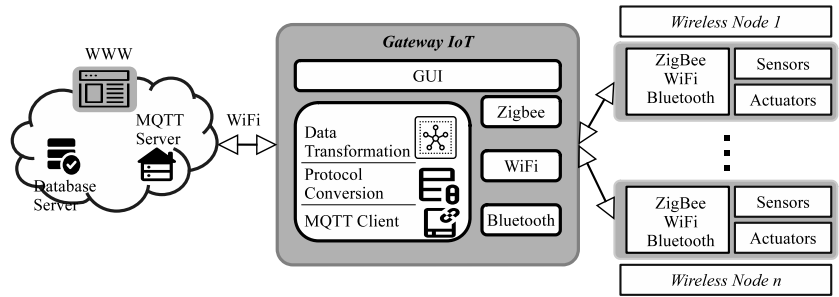}
\caption{Architecture of the developed IoT gateway}
\label{ch5_fig:3}       
\end{figure}

We have defined an architecture with five modules, which are summarized in Figure\ref{ch5_fig:3}. Each module is related to the main functions of the IoT gateway: (i) multiprotocol interconnection with remote nodes, (ii) data transformation, (iii) protocol conversion (iv) communication with IoT cloud platform and (v) user interface.

\begin{itemize}
\item Interconnection: the proposed architecture allows to interconnect several wireless nodes for data transmission, simultaneously, through different communication protocols (such as ZigBee, WiFi and Bluetooth). This communication is bidirectional, since the gateway collects data registered by wireless nodes through its sensors, but in addition, data can be sent to the nodes to operate on actuator devices.
\end{itemize}

\begin{figure}[ht!]
\centering
\includegraphics[scale=1.0]{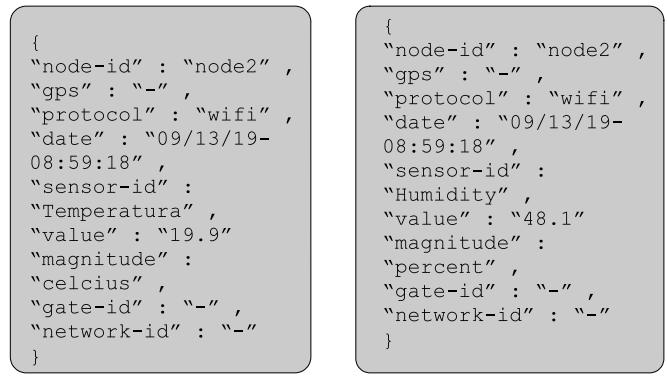}
\caption{Example of the standardized format for data transmission}
\label{ch5_fig:4}       
\end{figure}

\begin{itemize}
\item Data transformation: this module consists of standardization and establishment of a standard format to subsequently send data to a database server. This transformation is indispensable since data is sent from heterogeneous nodes that carried out the data capture through different types of sensors and that arrive through different network protocols. To standardize the representation of data, the JSON format was chosen, which has important advantages such as simplicity and low resource consumption. Also, a model is established for the identification of the devices that make up an IoT scenario. An example of the representation of the data recorded by the temperature and humidity sensor, in JSON format, is shown in Figure \ref{ch5_fig:4}. As can be seen in this figure, the first three fileds are "node-id", "gps" and "protocol", which correspond to the identification of the remote node, its gps coordinates and the protocol by which it transmitted the data. The following fields are "date", "sensor-id", "value" and "magnitude", which record the sensor information and its data. Finally, there is the “gate-id” and “network-id” fields, which will be used later to identify the gateway and the communication network. These last records will be useful when you have a scenario with more than one gateway and / or more than one transmission network.
\end{itemize}

\begin{figure}[ht!]
\centering
\includegraphics[scale=0.9]{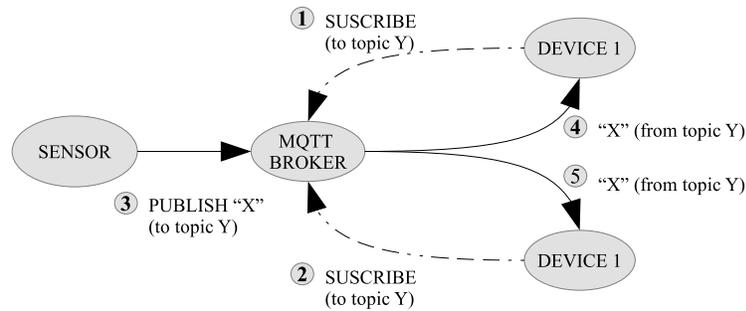}
\caption{Basic operation of MQTT protocol}
\label{ch5_fig:5}       
\end{figure}

\begin{itemize}
\item Protocol conversion: the proposed IoT gateway acts as a bridge or translator between different protocols, mainly between ZigBee, Bluetooth, WiFi and Ethernet. Therefore, the gateway continuously must be listening for connection requests on its network interfaces. The reception of data by any communication protocol involves the extraction of the payload and the subsequent construction of a new message with the standardized format. For data transmission from the gateway to database server in the Internet, MQTT (Message Queuing Telemetry Transport) \cite{ISO2016} protocol is used. This protocol is recommended in network scenarios in which bandwidth consumption must be reduced and where the devices involved in communication have low processing and memory capacity. The MQTT protocol works under a publication-subscription model that uses three components: a Broker, a Subscriber and a Publisher. A device can be registered as a Subscriber to a topic of interest in order to obtain information published on that topic. The Publisher is the generator of data for a specific topic. Data of a topic are transmitted to the Subscriber through the Broker. Hence, a Broker can be considered as a server that routes messages published to Subscribers (see Figure \ref{ch5_fig:5}).
\end{itemize}

\begin{itemize}
\item Communication with IoT cloud platform: this module consists of receiving data from the sensors already formatted and sending them to an MQTT broker hosted on the Internet (iot.eclipse.org). It is also responsible for receiving the configuration parameters that a user can introduce through the graphical interface and transmit them to remote nodes, also using MQTT protocol. In the first case, communication from the gateway to the broker, the gateway uses the Eclipse Paho Python Client library and acts as a publisher. In the second case, that is, during the transmission from the gateway to wireless nodes, the gateway operates as a publisher and the wireless nodes as subscribers. Because wireless nodes (sensor nodes) do not have an Internet connection, it is also necessary that the gateway has a broker installed for this type of message exchange. The broker installed in the gateway was the well-known broker Mosquitto \cite{Light2017}, which due to its characteristics of low resource consumption and reduced overload information, is suitable for being installed on embedded boards. 
\item User interface: finally, the fifth module of the gateway allows user to introduce some system configuration parameters, through a graphical interface. Also, allows users to register new nodes and/or sensors, modification of the frequency of data capture rates, the assignment of communication protocols, visualization of the measures registered in the database, the creation of rules for announcements and alarms, among other functions.
\end{itemize}

It is important to clarify that in the proposed model, the database where the data is finally stored and the web platform where data can be displayed, are hosted outside the gateway, specifically, on a private server on the Internet.

\section{A multiprotocol gateway for IoT: a proposed solution for the interoperability}
\label{sec:5.4}

\begin{figure}[ht!]
\centering
\includegraphics[scale=0.8]{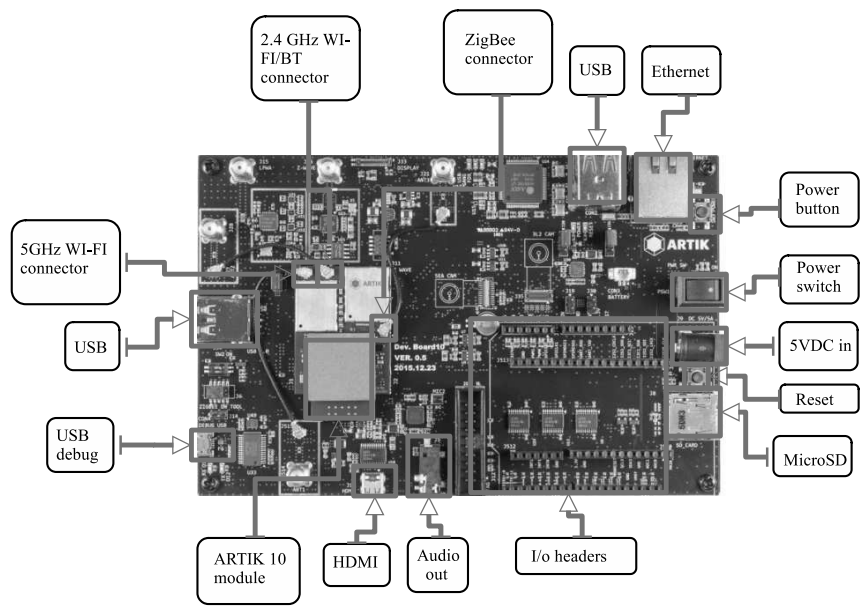}
\caption{Main technical specifications of the Artik 1020 board}
\label{ch5_fig:6}       
\end{figure}

For the development of the gateway, the Samsung Artik 1020 development kit was used, which is a high performance and multi-protocol embedded board that has the possibility of wireless communications through Bluetooth, ZigBee and WiFi. It also has multiple I/O ports capable of communicating through I2C, SPI, UART modules. A summary with the main technical specifications is shown in Figure \ref{ch5_fig:6}.

\section{A multiprotocol gateway for IoT: a proposed solution for the interoperability}
\label{sec:5.5}

\begin{figure}[ht!]
\centering
\includegraphics[scale=0.65]{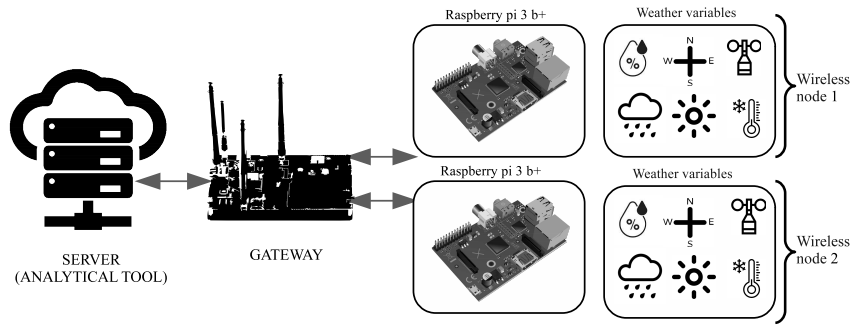}
\caption{Use Case implemented}
\label{ch5_fig:7}       
\end{figure}

In order to evaluate the developed gateway architecture, a use case was built. The implemented system involves an IoT gateway and two remote wireless nodes. Each wireless node has 6 sensors registering the following variables: environmental temperature, relative humidity, solar radiation, wind speed, rainfall level and wind direction. The wireless nodes transmit data registered by sensors to the gateway through ZigBee, WiFi and Bluetooth. This data transmission was configured as follows: the wind direction and wind speed variables were transmitted using Bluetooth; the radiation variables and rainfall level through ZigBee; and finally, temperature and humidity over WiFi. A schematic diagram summarizing the use case implemented is shown in Figure \ref{ch5_fig:7}.

The wireless nodes were implemented using the board Raspberry Pi 3 Model B with Ubuntu Mate as operating system. The sensors used were the following.

\begin{figure}[ht!]
\centering
\includegraphics[scale=1.0]{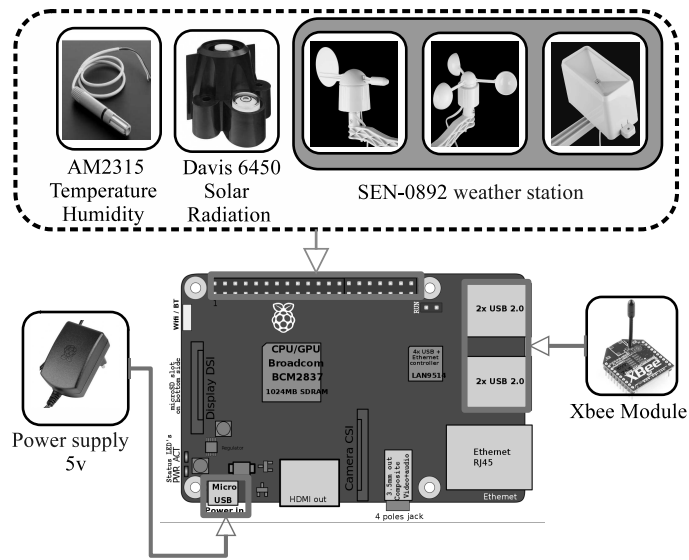}
\caption{Use Case implemented}
\label{ch5_fig:8}       
\end{figure}

The sensor used to measure the ambient temperature and relative humidity is the AM2315. The Davis 6450 sensor is responsible for measuring solar radiation. And finally, the SEN-08942 kit was used, which consists of an anemometer, a rain gauge and a wind vane to determine the wind direction. A summary with the main characteristics of the wireless nodes is shown in Figure \ref{ch5_fig:8}.

\begin{table}[]
\begin{tabular}{|l|p{3.4cm}|l|p{1.2cm}|p{1.2cm}|p{2.0cm}|}
\hline \hline
\textbf{Model} & \textbf{Microcontroller} 										 & \textbf{Frequency} & \textbf{RAM} & \textbf{FLASH} & \textbf{SD Support} \\ \hline \hline
 Artik 1020    & Quad core ARM Cortex A15, Quad core Cortex A7 & 1.5 - 1.3 GHz      &  2 GB				 & 16 GB					& 32 Gb 									 \\ \hline
 Raspberry pi  & Broadcom BCM2837B0, Cortex-A53 64-bit         & 1.4 GHz						&  1 GB				 & 16 GB 					& 64 Gb 									 \\ \hline \hline
\end{tabular}
\caption{Main features of Artik and Raspberry Pi B+ boards}
\label{ch5_tab:1}       
\end{table}

Tables \ref{ch5_tab:1} and \ref{ch5_tab:2} show a comparison between the two embedded cards that were used for the implementation of the use case.

\begin{table}[]
\begin{tabular}{|l|l|l|p{1.2cm}|p{1.2cm}|p{1.2cm}|p{1.2cm}|p{1.2cm}|}
\hline \hline
\textbf{Model} & \textbf{Analog input} & \textbf{Digital I/O} & \textbf{UART} & \textbf{SPI} & \textbf{I2C} & \textbf{PWM} & \textbf{USB} \\ \hline \hline
Artik 1020     & 6 										 & 95 							 		& Si  					& Si  				 &	4						& 2						 & Si  					\\ \hline
Raspberry pi   & 0 							       & 48	  						 		& Si 						& Si 					 &	Si					&	2						 & Si					  \\ \hline \hline
\end{tabular}
\caption{I/O and Buses of Artik and Raspberry Pi B+ boards}
\label{ch5_tab:2}       
\end{table}

\subsection{Sensors}

\begin{figure}[ht!]
\centering
\includegraphics[scale=0.2]{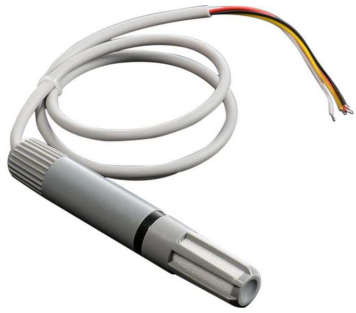}
\caption{AM2315 Sensor for environmental temperature and relative humidity}
\label{ch5_fig:9}       
\end{figure}

The sensor used to measure the environmental temperature and relative humidity was the AM2315 sensor. This has a humidity measurement accuracy of $\pm$ 2 \% and a temperature measurement of $\pm$ 0.1 C. Communication with the embedded system Raspberry pi is through the standard I2C protocol. In Figure \ref{ch5_fig:9} the mentioned sensor is shown. 

\begin{figure}[ht!]
\centering
\includegraphics[scale=0.15]{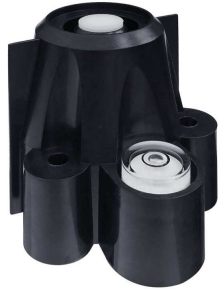}
\caption{Solar radiation sensor}
\label{ch5_fig:10}       
\end{figure}

On the other hand, for registration of the solar radiation, the Davis 6450 sensor was used. This transducer offers an analog output of 0 to 3Vdc, with a resolution of 1.67mV per W/m2 \cite{DavisInstruments2014}. In Figure \ref{ch5_fig:10} Davis 6450 sensor is exposed.

Finally, the measuring kit SEN-08942, which has three different types of sensors to measure rainfall level, direction and speed of the wind. The rain gauge operates by momentary closing a switch every 0.011 inches (0.2794 mm) of rainwater collected in a bucket. These contacts are recorded with a counter activated by an interrupt input. An anemometer is used to measure the wind speed, a switch commanded by an encoder is activated for each rotation of the blades. Finally, weather vane is a combination of resistances which are used for measuring the wind direction through a voltage division \cite{SparkFunElectronics2009}. The measuring kit SEN-08942 is set forth below in Figure \ref{ch5_fig:11}.

\begin{figure}[ht!]
\centering
\includegraphics[scale=0.5]{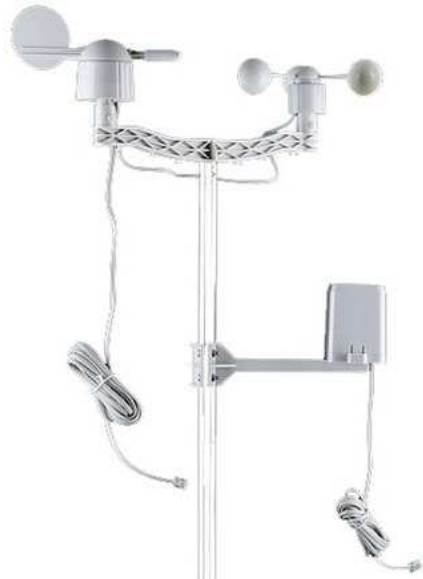}
\caption{Weather station kit, SEN-08942}
\label{ch5_fig:11}       
\end{figure}

\subsection{Data Analytic Tool}

\begin{figure}[ht!]
\centering
\includegraphics[scale=1.0]{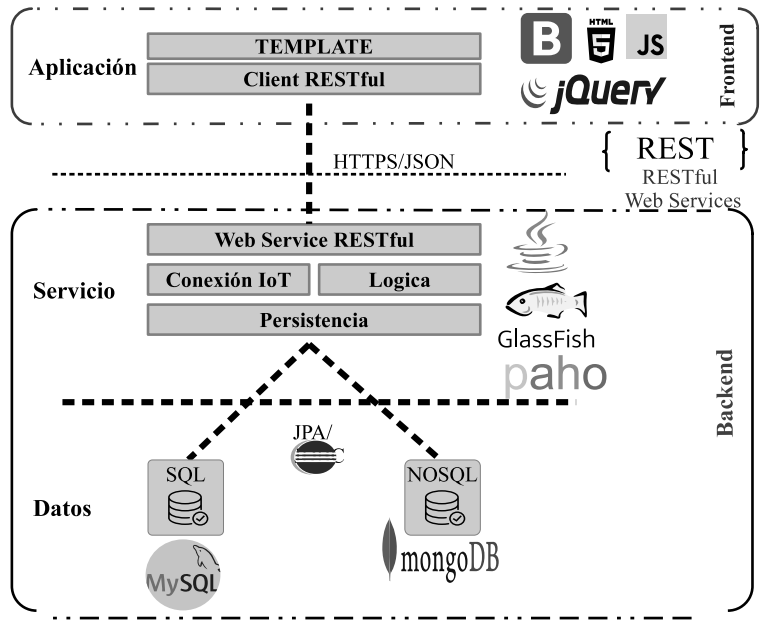}
\caption{Architecture of the data analytic tool}
\label{ch5_fig:12}       
\end{figure}

The data analytic tool was developed to support client/server and publisher/subscriber communication architectures with the IoT gateway. It has an architecture by decoupled layers, with RESTful Web Service for communication between Backend and Frontend (see Figure \ref{ch5_fig:12}). The Backend has connection to MySQL 5.7 and MongoDB 4.2 databases through DAO and JPA, developed in JavaEE 7 on a Glassfish 4.1 server. For the connection with the broker, PAHO Eclipse 1.4 is used which publishes a Web Service with OAuth2 security to be consumed by the Frontend. The Frontend consumes the service with OAuth2 Client and a security token, the information is presented with Bootstrap chart according to the established data and variables.

\subsection{Performance evaluation}

\begin{figure}[ht!]
\centering
\includegraphics[scale=0.65]{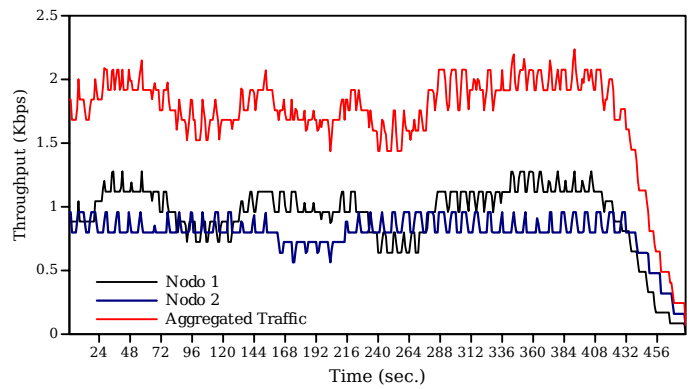}
\caption{Received Traffic over WiFi}
\label{ch5_fig:13}       
\end{figure}

\begin{figure}[ht!]
\centering
\includegraphics[scale=0.7]{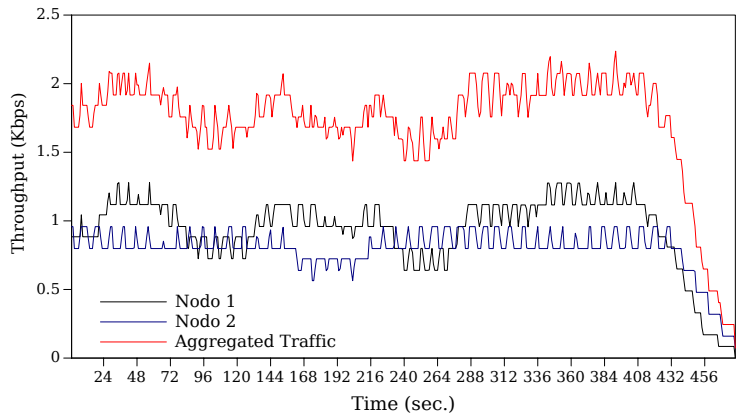}
\caption{Received Traffic over Bluetooth}
\label{ch5_fig:14}       
\end{figure}

The performance evaluation of the monitoring system was carried out by recording the traffic received from the two wireless nodes through different communication protocols. The system implemented, performed data capture and transmission to the gateway for 8 minutes.Figure \ref{ch5_fig:13} and \ref{ch5_fig:14} show the throughput of the traffic received by the WiFi and Bluetooth interfaces, respectively, for each nodes. For WiFi traffic, it is observed that the aggregated traffic does not exceed 2.5 Kbps. On average, the traffic throughput of node 1 is 1.1 Kbps and for node 2 of 1.3 Kbps. These values are very low, despite that transmission frequency of the recorded variables was 6 seconds. With this low level of traffic, we can consider that there is still enough bandwidth to support the interconnection of more remote nodes without impact the traffic flows. In terms of traffic over Bluetooth, node 1 had an average throughput of 0.95Kbps and node 2 approximately 0.85Kbps. This is equivalent to approximately 1.8Kbps on average in aggregate traffic. Based on these values, with Bluetooth traffic, as with WiFi traffic, they show that more nodes could be added by transmitting to the gateway through this protocol. 

\begin{figure}[hbt!]
\centering
\includegraphics[scale=0.65]{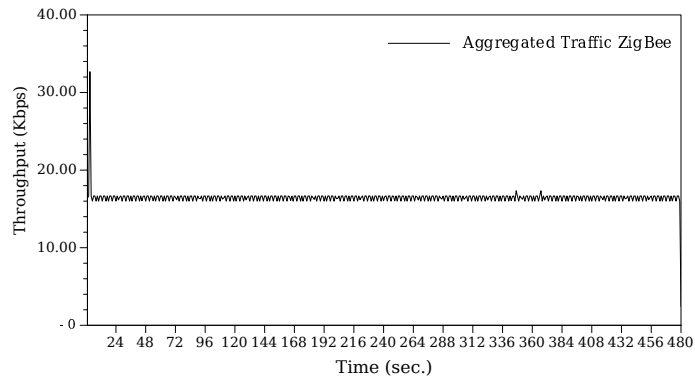}
\caption{Received Traffic over ZigBee}
\label{ch5_fig:15}       
\end{figure}

Similarly, Figure \ref{ch5_fig:15} shows the throughput achieve with ZigBee. The figure shows the aggregate traffic (node 1 + node2) and in this case, an average throughput of 16 Kbps was obtained. 

\begin{figure}[hbt!]
\centering
\includegraphics[scale=0.9]{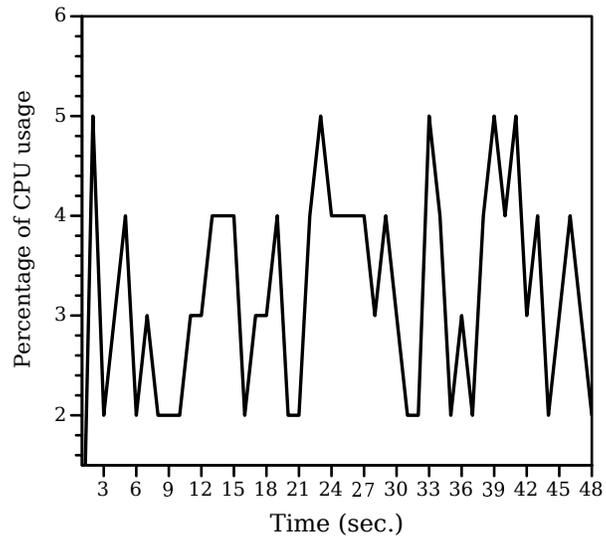}
\caption{Percentage of CPU usage}
\label{ch5_fig:16b}
\end{figure}

\begin{figure}[hbt!]
\centering
\includegraphics[scale=0.9]{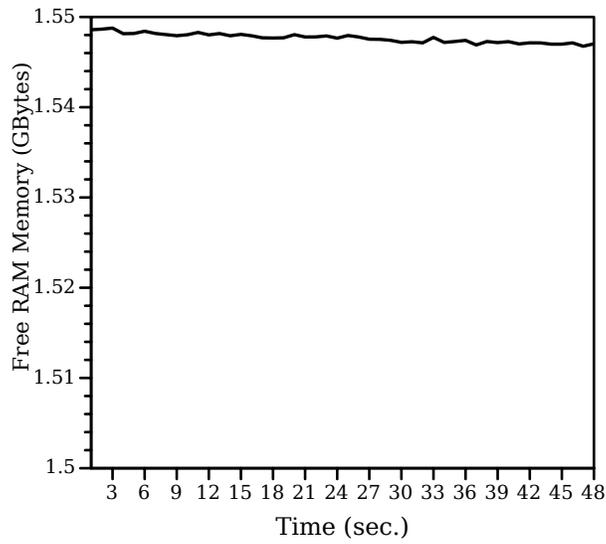}
\caption{Free RAM Memory}
\label{ch5_fig:16a}
\end{figure}

Finally, the percentage of CPU usage (see Figure \ref{ch5_fig:16b}) as well as the free memory were recorded (see Figure \ref{ch5_fig:16a}) every 10 seconds during system operation. That is, during the 8 minutes in which sensor nodes were transmitting environmental data to the gateway. Regarding CPU usage, the results show that only a 5\% of use was obtained. And in terms of RAM Memory, 1.5 GB of free memory was obtained on average, this represents an availability equivalent to 75\% of the total memory installed in the embedded board. The obtained results indicate that the processes implemented in the IoT gateway have a low consumption of resources, which would allow to incorporate more algorithms and more remote nodes simultaneously transmitting to the IoT gateway.

\subsection{Sensors Readings}

The following figures illustrate the sensors measurements for environmental temperature, relative humidity, wind speed and solar radiation. These variables were measured by wireless nodes for 7 minutes. Each figure has two curves which correspond to the measurements registered by the two nodes.

\begin{figure}[ht!]
\centering
\includegraphics[scale=1.0]{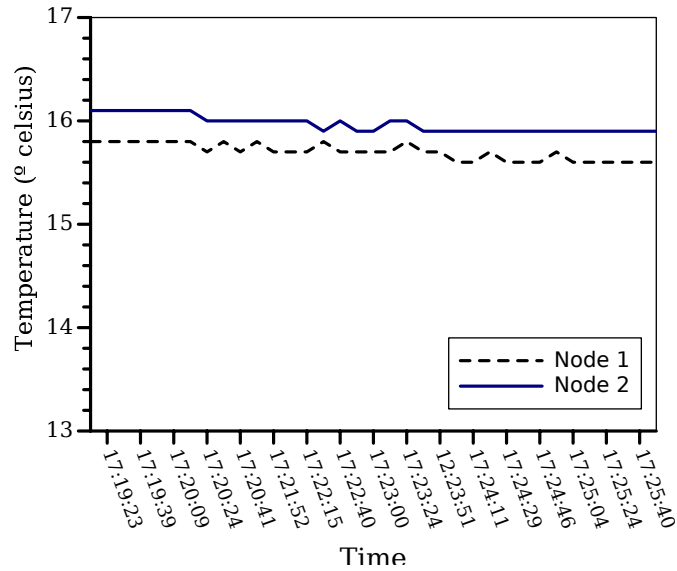}
\caption{Environmental temperature readings}
\label{ch5_fig:17}       
\end{figure}

As shown in Figure \ref{ch5_fig:17}, the difference in temperature readings is not significant due to wireless nodes were located close to each other. Therefore, only a difference of 0.3 Celsius was registered.

\begin{figure}[ht!]
\centering
\includegraphics[scale=1.0]{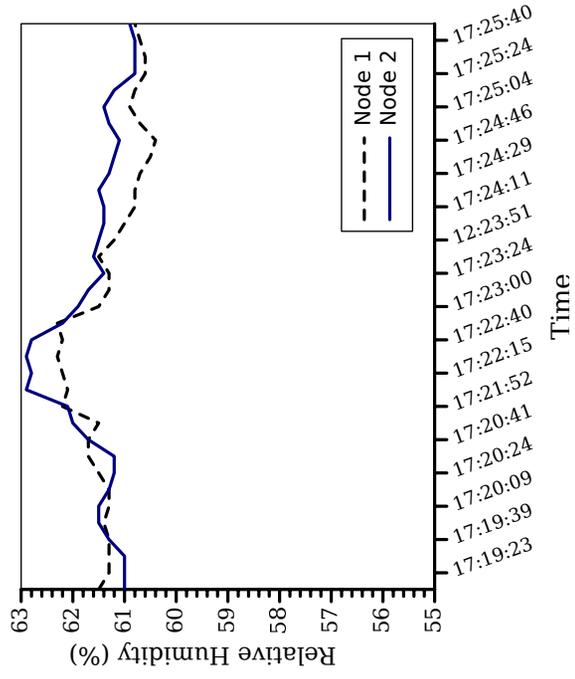}
\caption{Relative Humidity readings}
\label{ch5_fig:18}       
\end{figure}

Figure \ref{ch5_fig:18} Illustrates the relative humidity (in terms of percentage) measured by AM2315 sensors of both wireless nodes. The data variation is similar for the two nodes, obtaining a maximum difference in percentage of humidity of 1$\%$.

\begin{figure}[ht!]
\centering
\includegraphics[scale=1.0]{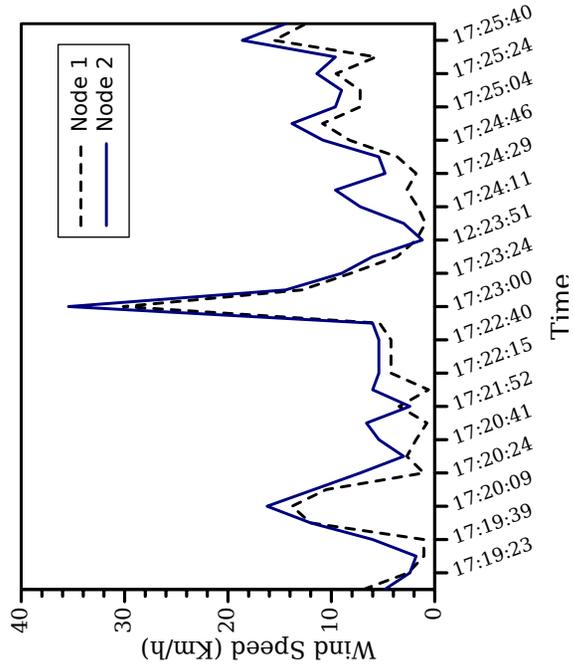}
\caption{Speed wind readings}
\label{ch5_fig:19}       
\end{figure}

Figure \ref{ch5_fig:19} shows the measurements of wind speed taken by the kit SEN-08942. Nodes were in different places obtaining the value of the wind speed in Km/h. Therefore, readings have a similar variation in the two nodes.

\begin{figure}[ht!]
\centering
\includegraphics[scale=1.0]{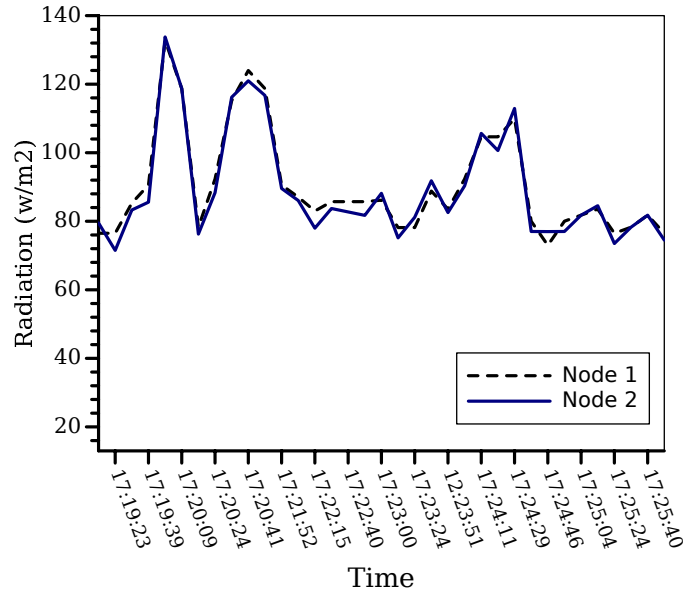}
\caption{Speed wind readings}
\label{ch5_fig:20}       
\end{figure}

Finally, Figure \ref{ch5_fig:20} shows the solar radiation variable captured by Davis 6450 sensor, data is generated in $W/m^{2}$.

\section{Conclusions}
\label{sec:5.7}

In this chapter, we describe the technical impact of IoT as well the impact over the world economy. A discussion about the heterogeneity of IoT devices and the problems that this fact generates was also exposed. As a possible solution to interoperability problem caused by the heterogeneity of IoT devices we proposed a IoT gateway. The implemented gateway acts as the central element in a model in which several wireless nodes can send data through different communication protocols, such as: WiFi, Bluetooth, ZigBee and Ethernet. Acceptable results were obtained in the performance evaluation of the gateway in terms of bandwidth consumption as well as the percentage of CPU and RAM memory usage.

One of the main advantages of the proposed gateway is that it allows to configurate remote wireless nodes and data transmission to a data analytics service hosted on an Internet server. This provides flexibility in data storage and visualization.

\section*{Acknowledgements}
 This work is supported by the Universidad de San Buenaventura under the project “PIICO: Plataforma para la Interoperabilidad de dispositivos del Internet de las Cosas heterog\'eneos” (FI 014-006)

\backmatter

\bibliographystyle{apacite}
\bibliography{
biblio_ch_5}
\printindex


\end{document}